\documentclass[12pt]{article}

\usepackage{amsfonts, amsmath, amssymb, amsthm}

\newcommand{\supp}{{\rm supp}\,}
\newcommand{\R}{{\mathbb R}}
\newcommand{\C}{{\mathbb C}}
\newcommand{\V}{{\mathbb V}}

\newcommand{\di}{{\rm d}}
\newcommand{\im}{{\rm Im\,}}

\newcommand{\maj}{w_{{\rm maj}}}
\newcommand{\mjz}{{\bf w}_{{\rm maj}}}

\newcommand{\wir}{w_{{\scriptscriptstyle IR}}}
\newcommand{\wuv}{w_{{\scriptscriptstyle UV}}}
\newcommand{\dv}[2]{\langle #1,#2\rangle}
\newcommand{\Wick}[1]{{:\!#1\!\!:}}

\begin{document}
 \begin{center}
 {\Large\bf
Wick Power Series in Indefinite Metric Field Theories
}

 \bigskip
 {\large A.~G.~Smirnov and M.~A.~Soloviev}
 \date{}

 \smallskip

 {\it  I.~E.Tamm Department of Theoretical Physics, P.~N.~Lebedev Physics
Institute, Leninsky prosp. 53, Moscow 117924, Russian Federation}
 \end{center}

\begin{abstract}
The analytic aspects of the operator realization of Wick power series of
infrared singular free fields are considered. Taking advantage of the
holomorphy properties of the two-point correlation function and its
Hilbert majorant in $x$-space, we solve in a general and model independent
way the problem of finding the adequate test function space on which a
given Wick series is convergent. Substantial attention is paid to the
proper formulation of the spectral condition in case the suitable test
functions are entire analytic in momentum space.
\end{abstract}

\section{Introduction}

This work is devoted to the functional--analytic aspects
of the operator realization of the Wick power series
\begin{equation}
\sum_k
d_k\Wick{\phi^k}(x)
\label{lab000}
\end{equation}
of a free field $\phi$
whose two-point vacuum expectation value $w(x_1-x_2)=
\dv{\Psi_0}{\phi(x_1)\phi(x_2)\Psi_0}$  does not satisfy the positivity
condition. The construction of such series exhibits some peculiarities
which have no analogues in the positive metric case and should be taken
into account in the covariant quantization of gauge theories. Of
particular interest are the problems of adequate choice of test
functions and of appropriate generalization of the spectral condition,
which have been raised by Moschella and Strocchi \cite{MoschellaStrocchi}.
As shown in \cite{Soloviev1, Soloviev2}, a simple way of overcoming these
problems is the use of a relevant Paley--Wiener--Schwartz--type theorem
for analytic functionals.  Here we outline the development
\cite{SmirnovSoloviev1, SmirnovSoloviev2} of this approach based on
exploiting in a systematic manner the analytic properties of the Hilbert
majorant of the two-point function $w(x_1-x_2)$.  We present a general and
model independent criterion which enables one to find easily the adequate
test functions on which a given Wick series is convergent.  Particular
emphasis is given to formulating the spectral condition for those quantum
fields that are well defined only under smearing with test functions
analytic in $p$-space and showing that it is satisfied for the sums of
Wick series with arbitrarily singular infrared behavior.

\section{Analytic Properties of the Hilbert Majorant}

We begin by recalling some of the standard facts concerning the indefinite
metric formalism \cite{MorchioStrocchi, BLOT}, restricting our
consideration to the case of a single neutral scalar field $\phi(x)$ in
$\tt d$-dimensional space-time (${\tt d}\geq 2$).
In this formalism, the state space $\cal H$ is endowed by a pseudo-Hilbert
structure. This means that $\cal H$ is provided with two sesquilinear forms
$\dv{\cdot}{\cdot}$ and $(\cdot,\cdot)$. Because of a failure in
positivity, the vacuum expectation values
$\dv{\Psi_0}{\phi(x_1)\ldots\phi(x_n)\Psi_0}$ are expressible in terms of
the indefinite metric $\dv{\cdot}{\cdot}$, whereas the positive scalar
product $(\cdot,\cdot)$ defines the notion of convergence in $\cal H$. The
forms
$\dv{\cdot}{\cdot}$ and $(\cdot,\cdot)$ are connected by the
relation
\begin{equation}
(\Phi,\eta\Psi)=\dv{\Phi}{\Psi},\quad \Phi,\Psi\in {\cal H},
\label{lab0}
\end{equation}
where $\eta$ is a self-adjoint involutory operator, which is called the
Krein operator. As a consequence,
\begin{equation}
|\dv{\Phi}{\Psi}|\leq \sqrt{(\Phi,\Phi)} \sqrt{(\Psi,\Psi)},\quad
\Phi,\Psi\in {\cal H}.
\label{lab00}
\end{equation}
\indent
Assuming that $\phi$ is a tempered
operator--valued distribution defined on Schwartz's test function space
$S$, we can formulate as usual all Wightman axioms with the exception of
the spectral condition. In the indefinite metric case, the implementers of
the translation group $U(\xi)$ are not unitary but pseudounitary and are,
as a rule, unbounded operators. For this reason, the
standard formulation of the spectral condition in terms of the generators
of translations becomes meaningless. However, one can express the spectral
condition in terms of the matrix elements of $U(\xi)$:
\begin{equation}
{\rm supp}\,\int\langle\Phi,\,U(\xi)\Psi\rangle\,e^{-ip\xi}\, {\rm
d}\xi\subset \Bar \V_+ ,
\label{lab3}
\end{equation}
where $\Bar \V_+$ is the closed upper light cone, and $\Phi,\Psi$ are
arbitrary vectors in the domain of $\phi$.
Equivalently, the
spectral condition can be expressed in terms of the support properties of
the vector-valued functional
\begin{equation}
\Psi(f)=\int\phi(x_1)\ldots\phi(x_n)\,f(x_1,\ldots,x_n)\,
{\rm d}x_1\ldots {\rm  d}x_n\Psi_0,
\label{lab3a}
\end{equation}
Namely, the inclusion
\begin{equation}
\supp \Hat \Psi \subset K_{n-}=\{(p_1,\ldots,p_n)\,:\,p_m+\ldots+p_n\in
\Bar \V_-,\quad \forall\,\, m=1,\ldots, n\}
\label{lab4}
\end{equation}
is valid, where $\Hat
\Psi$ is the Fourier transform of $\Psi(f)$.
This form of the spectral condition
can be used as long as
there are test functions of compact support in $p$-space. The case when
test functions in $p$-space are analytic is considered in Section 4.

The two-point function $w(x_1-x_2)$
serves as the kernel of the
sesquilinear form $\dv{f}{g}_S=\dv{\phi(f)\Psi_0}{\phi(g)\Psi_0}$ on
$S(\R^{\tt d})\times S(\R^{\tt d})$:
\begin{equation}
\dv{f}{g}_S = \int w(x_1-x_2)\Bar f(x_1) g(x_2)\,\di x_1 \di x_2,
\label{lab5a}
\end{equation}
while the form
$(f,g)_S=( \phi(f)\Psi_0,\phi(g)\Psi_0)$ defines the distribution $\maj$,
which is called the majorant of $w$:
\begin{equation}
(f,g)_S = \int \maj(x_1,x_2)\Bar f(x_1) g(x_2)\,\di x_1 \di x_2.
\label{lab5b}
\end{equation}
\indent
For a free scalar field $\phi$, the factorization property of the $n$-point
Wightman functions implies the relation
\begin{equation}
 \langle:\phi(f_1)\ldots\phi(f_m):\Psi_0,\,
:\phi(g_1)\ldots\phi(g_n):\Psi_0\rangle
 = \delta_{mn} \sum_\pi \prod_j
 \langle f_j, g_{\pi(j)}\rangle_S,
\label{lab6}
\end{equation}
where $f_i,g_j\in S$, $\pi$ runs over all permutations of indices, and $::$
denotes the Wick-ordered product defined in terms of Wightman functions by
the recursive relation
\begin{multline}
:\phi(f_1)\ldots\phi(f_n): = \phi(f_1):\phi(f_2)\ldots\phi(f_n): - \\
-\sum_{j=2}^n \dv{\Psi_0}{\phi(f_1)\phi(f_j)\Psi_0}_S
:\phi(f_2)\ldots\phi(f_{j-1})\phi(f_{j+1}) \ldots\phi(f_n):.
\nonumber
\end{multline}
A formula analogous to (\ref{lab6}) holds for the positive scalar
product:
\begin{equation}
 (:\phi(f_1)\ldots\phi(f_m):\Psi_0,\,
:\phi(g_1)\ldots\phi(g_n):\Psi_0)
 = \delta_{mn} \sum_\pi \prod_j
 (f_j, g_{\pi(j)})_S.
\label{lab7}
\end{equation}
It means that the state space ${\cal H}$ is obtainable by
the usual Fock procedure from the one-particle subspace
${\cal H}^1=\overline{\bigl\{\Psi: \Psi=\phi(f)\Psi_0,\,f\in S\bigr\}}$,
where bar
stands for the closure in the topology defined by $(\cdot,\cdot)$.
In particular,
the relation (\ref{lab7}) implies that ${\cal H}^1$ is mapped onto
itself by the
Krein operator $\eta$. Indeed, by (\ref{lab0}) and (\ref{lab6}),
the vector $\eta\phi(f)\Psi_0$ is $(\cdot,\cdot)$--orthogonal to all
vectors of the form $\Wick{\phi(g_1)\ldots\phi(g_m)}\Psi_0$ with $m\ne 1$.
Because of (\ref{lab7}) and the cyclicity of the vacuum, the linear span
of such vectors is dense in the orthogonal complement $({\cal
H}^1)^{\bot}$ and therefore $\eta\phi(f)\Psi_0$ lies in $({\cal
H}^1)^{\bot\bot}={\cal H}^1$. For a
continuous mapping, the image of the closure of a set is contained in the
closure of its image, whence $\eta {\cal H}^1\subset {\cal H}^1$.
Since $\eta$ is an involutory operator, this inclusion implies ${\cal
H}^1=\eta\eta{\cal H}^1\subset \eta {\cal H}^1$, and so $\eta{\cal
H}^1={\cal H}^1$.

The key observation, which is central to our approach and serves as
a starting point in \cite{SmirnovSoloviev1}, is that the
majorant inherits the momentum--space support properties of the
initial two-point function and, consequently, its analyticity properties in
$x$-space. The following lemma shows that the roots of this fact lie in
the general structure of the indefinite metric formalism.
\par\noindent{\bf Lemma.}
{\it Let $\phi(x)$ be a free field acting in the corresponding
Fock--Hilbert--Krein space. Then
\begin{equation}
\supp \hat \maj(p_1,p_2)\subset (\supp \Hat w)\times(-\supp\Hat w).
\nonumber
\end{equation}
}
{\it Proof. }The null space
$N_{<>}$ of the
form $\dv{\cdot}{\cdot}_S$ coincides with the null space $N_{()}$
of the form $(\cdot,\cdot)_S$.  Indeed, let $f\in N_{<>}$.
Then
$\dv{\phi(f)\Psi_0}{\phi(g)\Psi_0}=0$ for any $g\in S$ and, by continuity,
$\dv{\phi(f)\Psi_0}{\Phi}=(\phi(f)\Psi_0,\eta\Phi)=0$ for any $\Phi\in
{\cal H}^1$.  Since $\eta{\cal H}^1={\cal H}^1$, we see
that $f\in N_{()}$, and so $N_{<>}\subset N_{()}$.
Analogously, $N_{()}\subset N_{<>}$.

Now suppose the lemma is false and fix
a point $(q_1,q_2)\in \supp \hat \maj$ which
does not belong to $(\supp \Hat w)\times(-\supp\Hat w)$.
For definiteness, let $q_1\notin \supp\Hat w$.
To prove the statement,
it is sufficient to find a test function $f\in N_{<>}$
which does not belong to $N_{()}$. The relation
$\dv{f}{g}_S=\int \Hat w(p)\Bar{\Hat f}(p)\Hat g(p)\,\di p/(2\pi)^{\tt d}$,
which is obtained by rewriting
(\ref{lab5a}) in momentum space, implies that all test functions
such that $\supp \Hat f\cap \supp\Hat w=\varnothing$ belong to $N_{<>}$.
Let $O$ be an open neighbourhood of $q_1$ such that $O\cap \supp\Hat
w=\varnothing$. By
assumption, there exists a test function $H\in {\cal
D}(O\times\R^{\tt d})$
\footnote{
As usual, for an open set $O$ we denote by ${\cal D}(O)$ the
space composed of smooth functions whose support is contained in $O$.}
such that
$\hat \maj(H)\ne 0$.  Since ${\cal D}(O)\otimes {\cal D}(\R^{\tt d})$ is
dense in ${\cal D}(O\times\R^{\tt d})$, one can assume that
$H(p_1,p_2)=F(p_1)G(p_2)$, where $F\in {\cal D}(O)$ and $G\in
{\cal D}(\R^{\tt d})$. Taking into account the
relation (\ref{lab5b}) in the form $(f,g)_S=\int \hat\maj(p_1,p_2)\Bar{\Hat
f}(p_1)\Hat g(-p_2) \di p_1\di p_2/(2\pi)^{2\tt d}$
and setting $\Bar{\Hat f}(p)=F(p)$,
$\Hat g(-p)=G(p)$, we obtain that $(f,g)_S\ne 0$ and hence $f\notin
N_{()}$.  However, $\supp\Hat f\cap
\supp\Hat w=\varnothing$ and $f\in N_{<>}$.  The lemma is proved.

In view of the inclusion $\supp\Hat w\subset \Bar\V_+$, which is implied
by the spectral condition,
it immediately follows from the proved lemma
that $w_{{\rm maj}}$ is the boundary value of
an analytic function ${\bf w}_{{\rm maj}}(z, z')$ which is holomorphic in
the tubular domain $\{z,z':\, y= {\rm Im\,}z\in \V_-,\, y'= {\rm
Im\,}z'\in \V_+\}$. Moreover, for all $y\in \V_+$ the following bound holds
\begin{equation}
|{\bf w}(x-x'-2iy)|^2\leq |{\bf w}_{{\rm maj}}(x-iy,\, x+iy)|\,
|{\bf w}_{{\rm maj}}(x'-iy,\, x'+iy)|,
\label{lab8}
\end{equation}
where ${\bf w}(z)$ is the Wightman analytic function whose boundary value
is $w$. Indeed, taking $f(\xi)=(\nu/\sqrt\pi)^{\tt
d}e^{-\nu^2(\xi-x-iy)^2}$ and $g(\xi)=(\nu/\sqrt\pi)^{\tt
d}e^{-\nu^2(\xi-x'-iy)^2}$,
setting $\Phi=\phi(f)\Psi_0$, $\Psi=\phi(g)\Psi_0$, and
writing the left- and right-hand sides in
(\ref{lab00}) as integrals over a plane in the analyticity domain and
passing to the limit as $\nu\to\infty$, we immediately obtain
(\ref{lab8}).

\section{The Convergence Criterion for Wick Series}

When finding those test functions under smearing with which the Wick power
series is convergent, we make use
of the generalized Gelfand-Shilov
spaces of type $S$. Let us recall that the space $S_a^b$ in this class is
determined by two non-decreasing sequences of positive numbers $a_k$ and
$b_l$ and consists of smooth functions on ${\R}^n$ that satisfy the
bounds
$$
\sup_x \sup_{|\kappa|\leq k}\sup_{|\lambda|\leq l} |x^k\partial^l
f(x)|\leq CA^kB^la_kb_l,
$$
with constants $A,B,C$ depending on $f$. From
the functional analysis standpoint it is reasonable to impose the
conditions
$$
a^2_k\leq a_{k-1} a_{k+1},\quad a_{k+l}\leq
  C_1h_1^{k+l}a_ka_l,\qquad b^2_k\leq b_{k-1} b_{k+1},\quad b_{k+l}\leq
  C_2h^{k+l}_2b_kb_l,
$$
where $C_{1,2}$ and $h_{1,2}$ are
constants. The indicator functions
$a(r)=\sup_k r^k/a_k$, and
$b(s)=\sup_k s^k/b_k$
characterize the behavior of the test functions and of their
Fourier transforms at infinity, while in the context of QFT they indicate
the infrared and ultraviolet behavior of the fields defined on $S^b_a$.
The spaces defined by $a_k=k^{\alpha k}$, $b_l=l^{\beta l}$
$(\alpha,\beta\geq 0)$ are most often used in applications, but even a wider
framework is preferable for the problem under consideration because it
allows more precisely characterizing that behavior.

It is reasonable to require $d_1\ne 0$ and to construct the operator
realization not only of the series (\ref{lab000}) but also of all series
$\sum_k d'_k\Wick{\phi^k}(x)$ which are subordinate to the series
(\ref{lab000}) in the sense that $|d'_k|\leq C|d_k|$.
In this case,
the domain generated by applying the sums of
(\ref{lab000}) and all its subordinate series to the vacuum contains
the cyclic domain of the initial
field $\phi(x)=\sum_{k=0}^{\infty}\delta_{1,k}\Wick{\phi^k}(x)$
and therefore is
dense in $\cal H$.

The
convergence (in any sense) of (\ref{lab000}) and its subordinate series
smeared with test functions in $S_a^b$
implies that the repeated action of their sums on the vacuum
should be well defined. This means that for any $f_1,\ldots,f_n\in S_a^b$
the series
 \begin{equation}
 \sum\nolimits_{\{k_j\}}
 d_{k_1}^{(1)}\ldots d_{k_n}^{(n)} :\phi^{k_1}:(f_1)\ldots :\phi^{k_n}:(f_n)
 \Psi_0
\label{lab9}
\end{equation}
should be convergent in ${\cal H}$ provided
$|d_k^{(j)}|\leq C|d_k|$, $1\leq j\leq n$.
Conversely, as is shown in \cite{SmirnovSoloviev1},
if all series of the form (\ref{lab9})
are convergent, then (\ref{lab000}) and all its subordinate series averaged
with test functions in $S_a^b(\R^{\tt d})$ converge in the sense of a
strong graph limit (see
\cite{ReedSimon} for the definition), and their sums are well defined on a
common dense invariant domain.

It is easy to see that a series
$\sum_{\nu}\Psi_{\nu}$ of vectors in a Hilbert space is unconditionally
convergent provided the number series
$\sum_{\mu,\nu}(\Psi_{\mu},\Psi_{\nu})$ is absolutely convergent. Hence, to
establish the convergence of (\ref{lab9}), it is
sufficient to prove the convergence of
the series
\begin{equation}
  \sum_{k_1,\ldots,
 k_{2n}} \prod_{j=1}^{2n}|d_{k_j}|\,|(
\Wick{\phi^{k_1}}(f_1)\ldots\Wick{\phi^{k_n}}(f_n)\Psi_0, \,
\Wick{\phi^{k_{n+1}}}(f_1)\ldots\Wick{\phi^{k_{2n}}}(f_n)\Psi_0)|.
\label{lab10}
\end{equation}
Reducing the products of Wick monomials to the totally normally ordered
form by means of the Wick theorem and applying the Fock structure condition
(\ref{lab7}), we can express the scalar products in (\ref{lab10}) in terms
of the two-point function $w$ and its majorant $\maj$. It is also
convenient to pass in (\ref{lab10}) to the summation over the multi-indices
$K=\{k_{j,m},\,1\leq j<m\leq 2n\}$ with nonnegative integer components
which have the sense of the number of pairings between $\Wick{\phi^{k_j}}$
and $\Wick{\phi^{k_m}}$. As a result, the series (\ref{lab10}) takes the
form
\begin{equation}
\sum_KD_K\,|W^K(\bar f\otimes f)|,
\label{lab11}
\end{equation}
where $f=f_1\otimes\ldots\otimes f_n$ and
\begin{align}
W^K&=\prod_{1\le j<m\le n}w(x_m-x_j)^{k_{jm}}\prod_{
   n+1\le j<m\le 2n} w(x_j-x_m)^{k_{jm}}
\prod_{\genfrac{}{}{0pt}{}{1\le j\le n}{n+1\le m\le
  2n}}w_{{\rm maj}}(x_j,x_m)^{k_{jm}},&\nonumber\\
D_K&=\frac{\kappa!}{K!}\prod_{1\leq
  j\leq 2n} |d_{\kappa_j}|,\quad
  \kappa_j=k_{1,j}+\ldots+k_{j-1,j}+k_{j,j+1}+\ldots+k_{j,2n}.&
\nonumber
\end{align}
From the results of Section 2, it follows  that all distributions $W^K$ are
the boundary values of analytic functions ${\bf W}^K(z)$ which are
holomorphic in the same tubular domain. For this reason, we can study the
convergence of the series (\ref{lab11}) by means of the following theorem
proven in \cite{SmirnovSoloviev1}.
    \par \noindent {\bf Theorem 1.} {\it
Let $V$ be an open convex cone and let $(v_K)$ be a countable family of
tempered distributions which are the boundary values of functions ${\bf
v}_K(z)$ holomorphic in the tubular domain $T^V= \{z: {\rm Im}\,z \in V\}$.
If there exists a vector $\eta\in V$ such that $$ \sum_K \inf_{0<t<\delta}
    e^{st}\int\frac{|{\bf v}_K(x+it\eta)|}{a(|x|/A)}\, {\rm d}x\leq
    C_{\delta,\epsilon,A} b(\epsilon s) $$ for every positive $\delta$,
$\epsilon$ and $A$, then the family $(v_K)$ is unconditionally summable in
the space ${S'}_a^b$ which is dual of $S_a^b$.  }

We characterize
  the infrared and ultraviolet behavior of the majorant by a pair of
  monotone nonnegative functions
  $\wir$ and $\wuv$
  increasing as their arguments tend to infinity and to zero,
  respectively, and satisfying the bound
\begin{equation}
|\mjz(z, z')|\le C_1\,\wir(|z|+|z'|)+C_2\,\wuv(|y|+ |y'|),
\label{lab12}
\end{equation}
where $y=\im z$ and $y'=\im z'$ belong to the negative and positive
$y_0$-semi-axes, respectively.
On the coefficients $d_k$ of the series (\ref{lab000}) we impose the
restrictions
\begin{equation}
\lim_{k\to\infty} (k!|d_{2k}|)^{1/k}=0,\quad
|d_{k+l}|\le Ch^{k+l}|d_k|\,|d_l|.
\label{lab13}
\end{equation}
The first condition ensures that the Wightman functions of the sum of
(\ref{lab000}) are analytic in the usual domain of local QFT, whereas the
second one permits passing from multiple series over the multi-index $K$ to
one-tuple series over $|K|=\sum_{1\leq j<m\leq 2n} k_{jm}$.

The following convergence criterion is formulated in terms of the
characteristics $\wir$ and $\wuv$ and therefore solves the convergence
problem for Wick series in a model independent way.
\par\noindent {\bf Theorem 2.} {\it Let $\phi$ be a free field acting in the
pseudo-Hilbert space ${\cal H}$, and let the positive majorant of its
    correlation function satisfy the inequality $(\ref{lab12})$ in which
    $w_{{\scriptscriptstyle IR}}$ and $w_{{\scriptscriptstyle UV}}$
    are monotone. Under the conditions
    $(\ref{lab13})$ on the
    coefficients, the sums of the series $(\ref{lab000})$ and all its
    subordinate series are well defined as operator-valued generalized
    functions on every space $S^b_a$
    whose indicator functions satisfy the inequalities
\begin{equation}
\sum_k L^kk!\,|d_{2k}|\, w_{{\scriptscriptstyle IR}}(r)^k \le
      C_{L,\epsilon}a(\epsilon r), \quad
\inf_{t>0}e^{s t} \sum_k L^kk! |d_{2k}|\, w_{{\scriptscriptstyle
UV}}(t)^k \le C_{L,\epsilon} b(\epsilon s)
\nonumber
\end{equation}
for arbitrarily large $L>0$ and arbitrarily small $\epsilon>0$.}

The proof can be obtained by applying Theorem 1 to the
series of distributions $\sum_K D_K\,W^K(x)$ because the unconditional
convergence of this series in ${S'}_a^b$ is equivalent to the convergence
of (\ref{lab11}) for any $f\in S_a^b$. The details of the proof can be
found in \cite{SmirnovSoloviev1}. Here we only note that because of
(\ref{lab8}), the bound (\ref{lab12}) on the majorant proves to be
sufficient for estimating ${\bf W}^K(z)$, although the explicit expression
for $W^K$ contains not only $\maj$ but also the two-point function $w$.
\par\noindent
   {\bf Remark.} Since $\maj$ is a tempered distribution, its
   singularities are no worse than polynomial or
   logarithmic ones.  Therefore, it can be assumed that the inequalities
$$ w_{{\scriptscriptstyle IR}}(\lambda r)\leq C_\lambda
   w_{{\scriptscriptstyle  IR}}(r), \qquad w_{{\scriptscriptstyle
      UV}}(t/\lambda )\leq C'_\lambda w_{{\scriptscriptstyle UV}}(t), $$
hold for any $\lambda>0$, at least in the limiting sense, i.e., for
$r>R(\lambda)$ and $t<\delta(\lambda)$. In this case, the conditions
specified in Theorem 2
take a simpler form
\begin{equation}
\sum_k
L^kk!d_{2k} w_{{\scriptscriptstyle IR}}(r)^k \le C_{L}\,a(r),\quad
    \inf_{t>0}\,e^{st} \sum_k L^kk!  d_{2k} w_{{\scriptscriptstyle
      UV}}(t)^k \le C_{L}\, b(s),
\nonumber
\end{equation}
where $L>0$ is arbitrarily large. In the most important case of
the normal exponential $:\exp ig\phi:(x)$
we arrive at
$$
\exp\{L\,w_{{\scriptscriptstyle IR}}(r)\} \le
      C_{L}\,a(r), \quad \inf_{t>0}\,\exp\{st+L\,
      w_{{\scriptscriptstyle UV}}(t)\} \le C_{L}\, b(s).
$$

\section{Generalized Spectral Condition}

Simple models treated in \cite{MoschellaStrocchi} show that the local and
covariant formulation of gauge theories may require, for infrared reasons,
the use of test function spaces consisting of functions analytic in
$p$-space. The operator realization of the Wick exponential of the dipole
ghost field is a typical example.
In this case, the definitions (\ref{lab3}) and (\ref{lab4})
as well as the very notion of support of a generalized function break down.
In
\cite{Soloviev2, SmirnovSoloviev2} a generalization of the spectral
condition was proposed which is applicable to the case of arbitrarily high
infrared singularity.  In order to formulate this generalization, we need an
alternative description of test function spaces in terms of complex
variables. Namely, let us introduce the following definition.
\par\noindent {\bf Definition 1.} Let $\alpha(s)$ and $\beta(s)$ be
nonnegative continuous functions indefinitely increasing on the half-axis
$s\geq 0$, let $\alpha(s)$ be convex and differentiable for $s>0$, and let
$\beta(s)$ be convex with respect to $\ln s$ and satisfy the condition
$2\beta(s)\leq\beta(hs)$ with a constant $h>1$. We define ${\cal
E}_\beta^\alpha$ to be the inductive limit $\varinjlim_{A,B>0}{\cal
E}_{\beta,B}^{\alpha,A}$, where the Banach spaces ${\cal
E}_{\beta,B}^{\alpha,A}$ consist of entire analytic functions on $\C^n$
with the finite norm
$$
\|g\|_{A,B}=\sup_{p,q}|g(p+iq)|\exp\{-\alpha(A|q|)+\beta(|p|/B)\}.
$$
\indent From the listed restrictions on $\alpha$ and $\beta$,
it follows that the spaces ${\cal
E}_\beta^\alpha$ form a subclass of the spaces $S^a_b$.
Namely, ${\cal
E}_\beta^\alpha$ coincides with the space $S^a_b$ defined by
$a_k=\sup_{r\geq 0}r^k e^{-\alpha_*(r)}$ and $b_l=\sup_{s\geq 0}s^l
e^{-\beta(s)}$, where $\alpha_*(r)=\sup_{s>0}(rs-\alpha(s))$.
What is more, those restrictions ensure that for the generalized functions
defined on
${\cal E}_\beta^\alpha$, the notion of a carrier cone can be introduced
which replaces the usual notion of support, see \cite{SmirnovSoloviev2}
for proofs. Loosely speaking, a closed cone $K$ is called a carrier cone
 of a generalized function $u\in {{\cal E}'}_\beta^\alpha$ if $u$
decreases outside $K$ like $\exp\{-\alpha(|p|)\}$.
In a more formal language, this means that $u$ allows a
continuous extension to a test function space whose elements grow outside
$K$ as $\exp\{\alpha(|p|)$.
To be precise, with any open cone $U$ we associate the
space ${\cal E}_\beta^\alpha(U)$ which is defined exactly as ${\cal
E}_\beta^\alpha$ with the exception that the norm $\|g\|_{A,B}$ is
replaced by
$$
\|g\|_{U,A,B}=\sup_{p,q}|g(p+iq)|
\exp\{-\alpha(A|q|)-\alpha\circ \delta_U(Ap)+\beta(|p|/B)\},
$$
where
$\delta_U(p)$ is the distance from the point $p$ to the cone $U$.
\par\noindent {\bf Definition 2.} A closed cone $K$ is called a carrier
cone of $u\in {{\cal E}'}_\beta^\alpha$ if $u$ has
a continuous extension to the space ${\cal
E}_\beta^\alpha(K)=\varinjlim_{U\supset K\setminus\{0\}}
{\cal E}_\beta^\alpha(U)$.

Once we have the definition of carrier cones, the desired generalization
of the spectral condition becomes straightforward.
\par\noindent {\bf Definition 3.}
We say that a field $\phi(x)$ defined
on the test function
space ${\cal E}_\beta^\alpha$ satisfies the
generalized spectral condition if the Fourier transforms of the
vector--valued functionals (\ref{lab3a}) are carried by
the cones $K_{n-}$ defined in
(\ref{lab4}).

In \cite{SmirnovSoloviev2}, we have proved the following theorem showing
the relevance of this definition.
\par\noindent {\bf Theorem 3.} {\it Suppose the
assumptions of Theorem $2$ are fullfilled and furthermore the
space $S^a_b$ belongs to
the subclass of spaces described by Definition~$1$.
Then
the sums of the series $(\ref{lab000})$ and all its subordinate series
satisfy the generalized spectral condition.}

We expect that
the proposed description of spectral properties may be
useful for both
the analysis of concrete models with singular infrared behavior
and the self-consistent Euclidean formulation of general gauge
QFT.

\smallskip
\noindent{\bf Acknowledgements.} This work was supported
in part
by the Russian Foundation
for Basic Research Grant No. 99-01-00376 and
INTAS Grant No. 99-1-590 (A.~G.~S.), and in part by RFBR Grants No.
99-02-17916 (M.~A.~S.) and No. 00-15-96566.

\end{document}